\documentclass[aps, pra, 10pt, twocolumn, notitlepage, superscriptaddress]{revtex4-2}
\usepackage{natbib}
\usepackage{graphicx}
\usepackage{dcolumn} 
\usepackage{bm}
\usepackage{mathtools}
\usepackage{amssymb}
\usepackage{enumitem}
\usepackage[svgnames,dvipsnames]{xcolor}
\usepackage[normalem]{ulem}
\usepackage[mathlines]{lineno}
\usepackage[titletoc, title]{appendix}
\usepackage{bbold}
\usepackage{comment}
\usepackage{hyperref}
    \hypersetup{unicode=true,
        pdftoolbar=true,
        pdfmenubar=true,
        pdffitwindow=false,
        pdfstartview={FitH},
        pdfnewwindow=true,
        colorlinks=true,
        linkcolor=Maroon,
        citecolor=NavyBlue,
        filecolor=Maroon,
        urlcolor=NavyBlue}

\newcommand{\bra}[1]{\langle #1 |}
\newcommand{\ket}[1]{| #1 \rangle}
\newcommand{\braket}[2]{\langle #1 | #2 \rangle}
\newcommand{\braop}[3]{\langle #1 | #2 | #3 \rangle}

\newcommand{\beq}{\begin{equation}}
\newcommand{\eeq}{\end{equation}}
\newcommand{\beqa}{\begin{eqnarray}}
\newcommand{\eeqa}{\end{eqnarray}}
\newcommand{\ba}{\begin{array}}
\newcommand{\ea}{\end{array}}
\newcommand{\iunit}{\mathrm{i}}

\makeatletter
\newsavebox{\@brx}
\newcommand{\llangle}[1][]{\savebox{\@brx}{\(\m@th{#1\langle}\)}%
  \mathopen{\copy\@brx\kern-0.5\wd\@brx\usebox{\@brx}}}
\newcommand{\rrangle}[1][]{\savebox{\@brx}{\(\m@th{#1\rangle}\)}%
  \mathclose{\copy\@brx\kern-0.5\wd\@brx\usebox{\@brx}}}
\makeatother

\begin{document}

\title{Equilibrium and dynamical quantum phase transitions \\ in dipolar atomic Josephson junctions}
\author{Cesare Vianello}
\email{cesare.vianello@phd.unipd.it}
\affiliation{\mbox{Dipartimento di Fisica e Astronomia ``G. Galilei'', Università di Padova, Via F. Marzolo 8, I-35131 Padova, Italy}}
\affiliation{\mbox{Istituto Nazionale di Fisica Nucleare, Sezione di Padova, Via F. Marzolo 8, I-35131 Padova, Italy}}
\author{Giovanni Mazzarella}
\affiliation{\mbox{Liceo Scientifico ``G. B. Grassi'', Via B. Croce 1, I-21047 Saronno, Italy}}
\author{Luca Salasnich}
\affiliation{\mbox{Dipartimento di Fisica e Astronomia ``G. Galilei'', Università di Padova, Via F. Marzolo 8, I-35131 Padova, Italy}}
\affiliation{\mbox{Istituto Nazionale di Fisica Nucleare, Sezione di Padova, Via F. Marzolo 8, I-35131 Padova, Italy}}
\affiliation{\mbox{Padua QTech Center, Università di Padova, Via G. Gradenigo 6/A, I-35131 Padova, Italy}}

\begin{abstract} 

An atomic Josephson junction realized with dipolar bosons in a double-well potential can be described by an extended Bose-Hubbard model in which dipolar interactions generate an effective on-site interaction and nearest-neighbor pair tunneling. Using mean-field theory and exact diagonalization, we investigate how this correlated process affects zero-temperature equilibrium and dynamical properties of the system. In equilibrium, we show that pair tunneling induces ground-state parity modulations and significantly reshapes the phase diagram, producing qualitative changes in the quantum phase transitions toward NOON and phase-NOON states, as well as quantitative shifts of the critical points. Out of equilibrium, we demonstrate that it modifies the conditions for macroscopic quantum self-trapping, and assess its impact by comparing mean-field and fully quantum evolution, including the emergence of dynamical quantum phase transitions.

\end{abstract}

\maketitle

\section{Introduction}

The Josephson effect is one of the most relevant manifestations of macroscopic quantum coherence. Originally discovered in weakly linked superconductors (Josephson junctions) \cite{Josephson_1962, Anderson, Barone_book}, it was later recognized as a general phenomenon arising from coherent quantum tunneling, which could also be realized in coupled Bose-Einstein condensates \cite{Javanainen}. Owing to remarkable experimental advances in the control of ultracold atoms, Josephson dynamics has since become ubiquitous, emerging in systems ranging from superfluids to exciton polaritons and supersolids \cite{Pereverzev, Marchenkov_2000, Sukhatme, Cataliotti_2001, Albiez_2005, Levy2007, Lagoudakis, Abbarchi, Valtolina_2015, Spagnolli_2017, Burchianti, Kwon_2020, Luick, Kreil, Biagioni}. In the context of double-well-trapped Bose-Einstein condensates realizing a bosonic Josephson junction \cite{Albiez_2005, Levy2007}, interactions among tunneling atoms give rise to nonlinear phenomena absent in superconducting junctions, most notably macroscopic quantum self-trapping \cite{Smerzi, Raghavan}. For sufficiently strong interactions, the very nature of the tunneling process can be fundamentally altered. In the case of $N=2$ atoms, exact quantum-dynamical calculations have shown that increasing the interaction strength changes the dynamics from Rabi oscillations to correlated pair tunneling \cite{Schmelcher1, Schmelcher2}, a behavior that has also been observed experimentally \cite{Folling, Longhi}. In extended optical lattices, strong interactions likewise play a crucial role in driving transitions between different quantum phases \cite{Lewenstein_book}. These systems are typically described by extended Bose-Hubbard models \cite{Mazzarella-Giampaolo, Li, Guo, Zaleski}, which include---besides the standard single-particle tunneling and on-site two-body interaction terms---additional two-body operators accounting for nearest-neighbor density-density interactions, collisionally induced tunneling, and correlated pair tunneling. Although such processes can emerge from sufficiently strong contact interactions, they become particularly important in the presence of dipolar interactions, where inter-site interaction energies can compete with---and may even exceed---the on-site contributions \cite{Lewenstein, Maik}. 

In this work, we adopt the extended Bose-Hubbard framework to study the impact of dipolar interactions on the equilibrium and dynamical properties of an atomic Josephson junction, enabling us to access higher-order tunneling processes while remaining in the weakly-interacting regime. Because only two wells are involved, the model reduces to an effective two-site description with a renormalized on-site interaction and nearest-neighbor pair tunneling. The standard atomic Josephson junction is known to exhibit a quantum phase transition (QPT) toward a NOON state, corresponding to a fragmented condensate \cite{Cirac, Huang, Mazzarella, Trenkwalder, Vianello}. By combining mean-field theory and exact diagonalization, we show that the inclusion of pair tunneling qualitatively modifies this transition and promotes the emergence of a distinct QPT toward a phase-NOON state. We further explore the consequences for the Josephson dynamics, including macroscopic quantum self-trapping, and investigate the time-domain counterparts of quantum phase transitions, namely dynamical quantum phase transitions (DQPTs), both in the absence and in the presence of pair tunneling.

The paper is organized as follows. In Sec.~\ref{sec:model} we introduce the extended two-site Bose-Hubbard model and derive the corresponding mean-field theory. In Sec.~\ref{sec:gs} we investigate the ground-state properties of the system, showing that pair tunneling induces parity modulations in the ground-state Fock probabilities, modifies the QPT toward the NOON state---changing its nature from continuous to first order within a certain parameter range---and gives rise to a different continuous QPT toward a phase-NOON state. We determine the location of these transitions at the mean-field level by mapping out the zero-temperature phase diagram and confirm the analysis in the fully quantum setting through a finite-$N$ scaling study of the low-energy spectrum and the ground-state fidelity susceptibility. In Sec.~\ref{sec:dyn} we address the dynamical properties of the system. We compare the mean-field evolution with the quantum dynamics of atomic coherent states, determining in particular how pair tunneling modifies the mean-field conditions for macroscopic quantum self-trapping. We then analyze DQPTs, identified through finite-$N$ precursors of non-analyticities in the Loschmidt return rate and real-time zeros of the Loschmidt amplitude. Finally, in Sec.~\ref{sec:conclusion} we summarize our results and draw the conclusions. Appendix \ref{app:fig} presents a supplementary figure, while Appendices \ref{appA} and \ref{appB} provide some technical details on DQPTs.

\section{The model}\label{sec:model}

We consider an ultracold gas of $N$ dipolar bosons of mass $m$ polarized by an external electric or magnetic field along the $z$ direction and confined by an external potential $U_\text{ext}(\mathbf x)$. This is given by the superposition of a double-well trap in the axial direction $x$ and of an isotropic harmonic confinement with frequency $\omega_{\perp}$ in the transverse plane. In the presence of a transverse energy $\hbar \omega_{\perp}$ much larger than the characteristic energy of the bosons in the axial direction, our system can treated as quasi-one-dimensional. The boson-boson interaction derives from the sum of a contact potential and a long-range dipole-dipole potential, and is of the form ${V_\text{int}(\mathbf x-\mathbf x')=g\delta(\mathbf x-\mathbf x')+\gamma(\frac{1}{|\mathbf x-\mathbf x'|^3}-3\frac{|z-z'|^2}{|\mathbf x-\mathbf x'|^5})}$ \footnote{Here $g=4\pi\hbar^2 a_s/m$, with $a_s$ the $s$-wave scattering length, and $\gamma=\mu_0\mu^2/4\pi$ for magnetic dipoles ($\mu_0$ is the vacuum permeability and $\mu$ is the magnetic dipole moment) or $\gamma=d^2/4\pi \varepsilon_0$ for electric dipoles ($\varepsilon_0$ is the vacuum permittivity and $d$ is the electric dipole moment).}. In a two-mode approximation, the bosonic field operator is $\hat \Psi(\mathbf x)=\Phi_L(\mathbf x)\hat a_L + \Phi_R(\mathbf x)\hat a_R$, where $\Phi_{L(R)}(\mathbf x)$ are the condensate wavefunctions and $\hat a_{L(R)}$ are the annihilation operators in each well \cite{Ananikian}. Substituting this expression into the many-body Hamiltonian $\hat H = \int d^3\mathbf x\,\hat\Psi^\dag(\mathbf x)[-\frac{\hbar^2}{2m}\nabla^2+U_\text{ext}(\mathbf x)]\hat\Psi(\mathbf x) + \frac{1}{2}\int d^3\mathbf x\,d^3\mathbf x'\times \hat\Psi^\dag(\mathbf x)\hat\Psi^\dag(\mathbf x')V_\text{int}(\mathbf x-\mathbf x')\hat\Psi(\mathbf x')\hat\Psi(\mathbf x)$, one obtains an extended two-site Bose-Hubbard model \cite{Lewenstein, Maik, Li, Pizzardo}
\begin{align}\label{eBH}
    \hat H = &-J_0(\hat a^\dag_L\hat a_R + \text{h.c.})+\frac{U_0}{2}\left[\hat n_L(\hat n_L-1)+\hat n_R(\hat n_R-1)\right] \nonumber\\
    &+V\hat n_L\hat n_R-T[\hat a^\dag_L(\hat n_L+\hat n_R)\hat a_R + \text{h.c.}]\nonumber\\
    &+P(\hat a_L^\dag \hat a_L^\dag \hat a_R \hat a_R + \text{h.c.}),
\,\end{align}
where $\hat n_{L(R)}\equiv \hat a^\dag_{L(R)}\hat a_{L(R)}$ and h.c.~denotes the Hermitian conjugate. The parameters $J_0$ and $U_0$ represent the single-particle tunneling amplitude and the on-site interaction energy, respectively. The former originates from the one-body term of the many-body Hamiltonian, while the two-body term is responsible for the on-site interaction $U_0$, the nearest-neighbor density-density interaction $V$, the collisionally induced tunneling $T$, and the pair tunneling $P$. In the absence of pair tunneling ($P=0$), the effects of the $V$ and $T$ terms on the ground-state and dynamical properties of few bosons have been studied analytically in Refs.~\cite{Mazzarella-DellAnna, Dutta, Pizzardo}.

The total number of particles $\hat N \equiv \hat n_L+\hat n_R$ is conserved. As a consequence, the $T$ term simply renormalizes the tunneling energy, while the $V$ term renormalizes the on-site energy. Therefore the Hamiltonian \eqref{eBH} can be rewritten, up to constants, as
\begin{align}\label{Hamiltonian}
    \hat H = &-J(\hat a_L^\dag \hat a_R+\text{h.c.})+\frac{U}{2}(\hat n_L^2+\hat n_R^2)\nonumber\\
    &+P(\hat a_L^\dag \hat a_L^\dag \hat a_R \hat a_R + \text{h.c.}),
\end{align}
where $J \equiv J_0+T(N-1)$ and $U\equiv U_0-V$. Throughout this work we assume $J>0$ and $P>0$, while $U$ may take either sign, corresponding to repulsive (${U>0}$) or attractive ($U<0$) interactions. In particular, the presence of the density-density interaction $V$ allows the effective interaction $U$ to become negative even when the bare on-site term $U_0$ is positive, thereby circumventing the instability associated with the collapse of the condensate cloud \cite{Mazzarella-DellAnna}.

Our analyses will be based on the numerical solution of the eigenvalue problem $\hat H \ket{E_n}=E_n \ket{E_n}$ for a fixed number $N$ of bosons, obtained by exact diagonalization (see e.g.~Refs.~\cite{Mazzarella, Mazzarella-DellAnna, Vianello}). In this case the Hilbert space is $(N+1)$-dimensional and the Hamiltonian is represented in the Fock basis ${\ket{i,N-i}} =\ket{i}_L \otimes \ket{N-i}_R$ $(i=0,\dots,N)$ by a $(N+1)\times (N+1)$ real symmetric matrix. Hence each eigenvalue $E_n$ is associated to an eigenstate $\ket{E_n}$ of the form
\begin{equation}\label{eigenstate}
    \ket{E_n} = \sum_{i=0}^N c_i^{(n)}\ket{i,N-i},
\end{equation}
with real coefficients $c_i^{(n)}$ normalized to unity.

The quantum dynamics generated by the Hamiltonian \eqref{Hamiltonian} has a well-defined semiclassical (mean-field) limit. This can be derived from the principle of least action for $S[\{q\}] =\int dt\,\braop{\Psi}{\iunit\hbar\partial_t-\hat H}{\Psi}$, where $\ket{\Psi}\equiv \ket{\Psi(\{q\})}$ is a trial macroscopic state constructed to encode as much information as possible about the underlying microscopic dynamics and parametrized by a set of classical variables $\{q\}$ \cite{Amico}. For the two-site Bose-Hubbard model, the standard choice is a Glauber coherent state $\ket{\bm a}\equiv\ket{a_L}\otimes \ket{a_R}$ \cite{Glauber, Zhang}, which can be parametrized by the relative phase $\phi\equiv\phi_R-\phi_L$ between the condensates in the two wells and the fractional population imbalance $z\equiv (n_L-n_R)/N$. The action is then
\begin{equation}\label{action}
    S(\phi,z) \equiv N s(\phi,z)=N \int dt\left[\frac{\hbar z}{2}\dot\phi - J\mathcal E(\phi,z)\right],
\end{equation}
where
\begin{align}\label{energy}
    \mathcal E &= - \sqrt{1-z^2}\cos\phi + \frac{\Lambda}{2}z^2 +\frac{\Pi}{2}(1-z^2)\cos 2\phi
\end{align}
is the classical energy in units of $J$, and 
\begin{equation}
    \Lambda \equiv \frac{UN}{2J}, \qquad \Pi \equiv \frac{PN}{J}\ge0.
\end{equation}
Imposing $\delta S=0$, we obtain the equations of motion
\begin{subequations}\label{semiclas}
    \begin{align}
        \hbar\dot\phi &= \frac{2Jz}{\sqrt{1-z^2}}\cos\phi + UNz - 2PNz \cos 2\phi,\\
        \hbar\dot z &= -2J \sqrt{1-z^2}\sin\phi + 2PN(1-z^2)\sin 2\phi.
    \end{align}
\end{subequations}
The same equations can be derived from the two-mode approximation of the Gross-Pitaevskii equation \cite{Mazzarella-Moratti_2009, Abad}, and in the absence of pair tunneling they reduce to the familiar equations for an atomic Josephson junction \cite{Smerzi, Raghavan}. Up to a reinterpretation of the parameter $P$, they also provide an improved semiclassical description of the standard (non-polarized) junction, accounting for higher-order overlaps between $\Phi_L(\mathbf x)$ and $\Phi_R(\mathbf x)$ \cite{Ananikian, Anglin, Mazzarella-Moratti_2010, Mele-Messeguer}. 

\section{Ground state}\label{sec:gs}

We begin by analyzing the effects of pair tunneling on the ground state $\ket{E_0}= \sum_{i=0}^N c_i^{(0)}\ket{i,N-i}$. In doing so, it is important to account for the fact that the $V$, $T$, and $P$ terms in Eq.~\eqref{eBH} originate from the same microscopic interaction and thus are not independent. In extended optical lattices, the typical hierarchy is such that $T$ is approximately one order of magnitude smaller than $V$, while $P$ is further suppressed by an additional order of magnitude and is therefore often neglected \cite{Lewenstein, Maik, Lewenstein_book}. For a single double-well system, however, the pair-tunneling amplitude $P$ can be comparable to $T$, depending on the barrier height and inter-well separation \cite{Pizzardo}. Motivated by this, we consider an illustrative protocol in which a system with $N=32$ particles is initially prepared in the ground state for $U_0/J>0$, and the dipolar interaction is then adiabatically increased from zero. As $V$ grows, we constrain $P$ and $T$ to scale as $P=T=0.1V$. Under this protocol, $U/J$ decreases while $P/J$ simultaneously increases. Since the particle number is moderate, we may safely assume that the bare tunneling amplitude dominates over interaction-induced corrections, i.e. $J_0 \gg T(N-1)=0.1(N-1)V$, so that the effective tunneling can be approximated as $J \simeq J_0$. In Fig.~\ref{fig1}, we compare this scenario with that in which dipolar interactions are absent and the same variation of $U/J$ is achieved by adiabatically tuning the contact interaction.

\subsection{Parity modulations}\label{sec:pm}

\begin{figure}
    \centering
    \includegraphics[width=\linewidth]{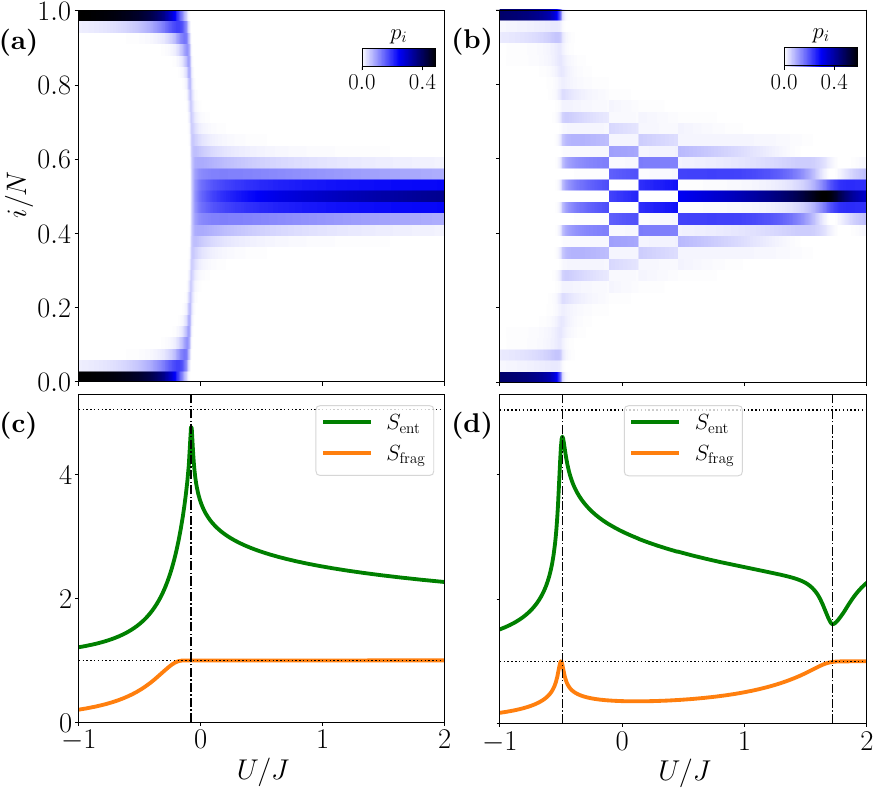}
    \caption{\textbf{Ground state.} Distributions of the ground-state probabilities $p_i$ [panels (a)-(b)] and the corresponding entanglement and fragmentation entropies [panels (c)-(d)] as functions of $U/J$, for $N=32$, $U_0 = 2.0$, $J=1.0$, and $V$ varying from 0 to 3.0, i.e. $U=U_0-V$ varying from $2.0$ to $-1.0$. The pair tunneling is $P = 0$ in panels (a) and (c), and $P = 0.1V$ in panels (b) and (d). Horizontal dotted lines indicate $\max S_\text{frag}=1.0$ and $\max S_\text{ent}=\log_2 33\simeq 5.04$, while vertical dashed-dotted lines mark local extrema of $S_\text{ent}$.}
    \label{fig1}
\end{figure}

We observe that even weak pair tunneling produces a clear signature in the distribution of the ground-state probabilities $p_i \equiv |c_i^{(0)}|^2$, appearing as a pronounced modulation of $p_i$ according to the parity of the index $i$ [Fig.~\ref{fig1}(a)-(b)]. The origin of this effect is that in the limit $J\to0$, the Hamiltonian contains only operators that change $n_{L(R)}$ by zero or two units. Consequently, the parity of $n_{L(R)}$ is conserved. In this limit, the Hilbert space decomposes into the direct sum of two disconnected sectors (fragments), $\mathcal H_\text{even(odd)}=\text{span}\{\ket{i, N-i}:i=0(1),2(3),4(5),\dots\}$, with $\dim \mathcal H_\text{even}= \lfloor N/2 \rfloor +1$ and $\dim \mathcal H_\text{odd}= \lceil N/2 \rceil$. In particular, the ground state has well-defined parity, belonging either to $\mathcal H_\text{even}$ or $\mathcal H_\text{odd}$. For $J\neq 0$, $\mathcal H_\text{even}$ and $\mathcal H_\text{odd}$ are coupled and the parity symmetry is explicitly broken, yet a parity-dependent modulation of the ground-state probabilities persists, and is enhanced with increasing $P/J$. To quantify the degree to which the ground state is hybridized across the two parity sectors, we introduce the projectors $\Pi_\text{even(odd)} = \sum_{i \text{ even(odd)}}\ket{i, N-i}\bra{i, N-i}$ and the corresponding weights $p_\text{even(odd)} = \braop{E_0}{\Pi_{\text{even(odd)}}}{E_0}$. We then characterize the resulting degree of fragmentation through the entropy
\begin{equation}
\label{frag}
    S_\text{frag}=-p_\text{even}\log_2p_\text{even} - p_\text{odd}\log_2p_\text{odd},
\end{equation}
taking values in the interval $[0, 1]$. Values of $S_\text{frag} < 1$ indicate that the two symmetry sectors are not equally weighted in the ground state, and $S_\text{frag}=0$ implies that the ground state has well-defined parity. Indeed, the probabilities modulation displayed in Fig.~\ref{fig1}(b) is associated to the reduction of $S_\text{frag}$ shown in Fig.~\ref{fig1}(d). In the absence of pair tunneling, $S_\text{frag}$ is generally equal to one, except when $N$ is even and $U/J$ is reduced below the critical value associated with the crossover toward the NOON state $\ket{N,0}+\ket{0,N}$. In fact, if $N$ is even the NOON state belongs to $\mathcal H_\text{even}$ and the corresponding fragmentation entropy is zero. This behavior, visible in Fig.~\ref{fig1}(c), is clearly absent when $N$ is odd, because then the NOON state does not have definite parity.

Prior to the crossover toward the NOON state, as $P/J$ increases there exist specific values of the interaction at which sudden changes in the parity of the probabilities modulation occur [Fig.~\ref{fig1}(b)]. These events correspond to level crossings of the two lowest eigenvalues, $E_0$ and $E_1$, and the observed parity switches result from the fact that $\ket{E_0}$ and $\ket{E_1}$ have probabilities modulations of opposite parity. These level crossings already occur for $N=4$ and $N=6$, for which there are two crossings \cite{Links}. For $N = 8$ there are three crossings, and for $N>8$ the number of crossings increases by one each time two particles are added to the system. Under our driving protocol, the values of $U/J$ corresponding to the first two (lowest $U/J$) crossings are nearly independent of $N$, while the position of last crossing shifts to larger $U/J$ (smaller $P/J$) with increasing $N$. Between the second and the last crossing there are $(N-8)/2$ additional crossings and hence further parity changes. However, the gap $E_1-E_0$ in the non-zero regions between the second and the last crossing decreases by roughly one order of magnitude with every two additional particles. Consequently, for large $N$ the two lowest eigenvalues become quasi-degenerate throughout this interval and are exactly degenerate at a dense set of points. For this reason the resolution in Fig.~\ref{fig1}(b) is insufficient to resolve the 12 parity changes that occur within the quasi-degenerate region for $N=32$. In the thermodynamic limit $N \to \infty$, the quasi-degenerate region becomes fully degenerate; the smallest value of $P/J$ at which a level crossing occurs then defines the critical point of a continuous QPT induced by pair tunneling. This transition will be analyzed in detail in Sec.~\ref{sec:qpt}.

From an experimental standpoint, the fact that $\ket{E_0}$ and $\ket{E_1}$ exhibit complementary parity modulations implies that this feature can be easily washed out in experiments unless $k_BT \ll E_1-E_0$. Nevertheless, other features are robust against finite temperature. For instance, at moderate pair tunneling the probabilities distribution is broader [Fig.~\ref{fig1}(a)-(b)], making imbalanced configurations more likely to be observed than in the absence of pair tunneling at the same temperature. Another effect expected to be robust against finite temperature is the shift of the critical interaction associated with the transition toward the NOON state \cite{Vianello}, which is discussed in the next section.

\subsection{Quantum phase transitions}\label{sec:qpt}

Besides parity properties, the spreading of the ground-state probabilities $p_i$ across the Fock states can be quantified by the entanglement entropy \footnote{This is the von Neumann entropy of the reduced density matrix $\hat\rho_R=\sum_{i=0}^{N} p_i \ket{i, N-i}\bra{i,N-i}$, obtained as the partial trace of $\hat{\rho}=\ket{E_0}\bra{E_0}$ over either the left or the right well, and therefore is a measure of the bipartite entanglement between the two wells \cite{Mazzarella, Vianello}.}
\begin{equation}
\label{ent}
    S_\text{ent}=-\sum_{i=0}^N p_i \log_2 p_i,
\end{equation}
which takes values in the interval $[0,\,\log_2(N+1)]$, with larger values of $S_\text{ent}$ indicating a more delocalized distribution. In particular, the critical interaction strength $U_c/J$ associated with the crossover toward the NOON state corresponds, up to $\mathcal O(N^{-1})$ corrections, to a local maximum of $S_\text{ent}$ [Figs.~\ref{fig1}(c)-(d)] \cite{Mazzarella, Vianello}. The clearest effect of pair tunneling is a shift of $U_c/J$; with the parameters used in Fig.~\ref{fig1}, this moves $U_c/J$ toward more negative values, namely stronger attractive interactions, with respect to the $P=0$ case. Importantly, beyond shifting the critical interaction strength, pair tunneling may qualitatively change the nature of the transition. For absent or weak pair tunneling, the crossover is accompanied by a closing of the gap $E_1-E_0$ in the thermodynamic limit, leading to a continuous QPT. By contrast, in the presence of sufficiently strong pair tunneling, the gap at $U_c/J$ develops a cusp, indicative of a sharp avoided level crossing, which in the thermodynamic limit becomes a level crossing giving rise to a first-order QPT.

The different nature of these transitions---as well as the pair-tunneling-induced transition anticipated toward the end of Sec.~\ref{sec:pm}---can be clearly captured at the mean-field level. We will later substantiate the mean-field analysis through a numerical study of the $N$-scaling of the energy gap and the ground-state fidelity susceptibility; see Fig.~\ref{fig3} and the accompanying discussion. The critical value $\Lambda_c \equiv U_c N/2J$ can be predicted from the bifurcation of the minimum of the classical energy $\mathcal E(\phi, z)$ [Eq.~\eqref{energy}] in the direction of $z$. The dependence of $\Lambda_c$ on $\Pi$ is piecewise. For $\Pi < \frac{1}{2}$, that we identify with the weak pair tunneling regime,
\begin{equation}
    \Lambda_c=-1+\Pi.
\end{equation}
In this regime, for $\Lambda > \Lambda_c$ there is a single minimum at $(\phi,z)=(0,0)$, while for $\Lambda <\Lambda_c$ there are two degenerate minima at $(0,\pm z_s)$, where
\begin{equation}
    z_s \equiv \sqrt{1-(\Lambda-\Pi)^{-2}}.
\end{equation}
At the critical point $\Lambda = \Lambda_c$, the minimum at $(0,0)$ undergoes a pitchfork bifurcation: $\partial^2_z\mathcal E(0,0)=0$, while $\partial^2_\phi\mathcal E(0,0)>0$. As $\Lambda$ is further reduced, the new imbalanced minima at $(0, \pm z_s)$ emerge continuously from it, while $(0,0)$ becomes a saddle. This represents a continuous QPT in terms of the order parameter $z$, and the closing of the gap $E_1-E_0$ at the critical point is a direct consequence of the softening of the collective mode along the direction of $z$. This is qualitatively the same behavior observed for $\Pi = 0$.

By contrast, for strong pair tunneling, $\Pi > \frac{1}{2}$, the critical interaction is
\begin{equation}
    \Lambda_c=-\Pi.
\end{equation}
For $\Lambda > \Lambda_c$ there are two degenerate minima at $(\pm \phi_s, 0)$, where
\begin{equation}
    \phi_s \equiv\text{arcsec}\left(2\Pi\right),
\end{equation}
while for $\Lambda < \Lambda_c$ there are again two degenerate imbalance minima at $(0, \pm z_s)$. At $\Lambda=\Lambda_c$, the two classes of stationary points coexist with the same energies, $\mathcal E(\pm \phi_s,0)=\mathcal E(0, \pm z_s)$, and upon crossing $\Lambda_c$ they exchange stability: $(0, \pm z_s)$ are saddles for $\Lambda>\Lambda_c$, while $(\pm \phi_s, 0)$ become saddles for $\Lambda < \Lambda_c$. This exchange of stability between two disconnected semiclassical minima leads to a first-order QPT. Since the phase and imbalance minima exist at finite distance in phase space, no mode softens, and the excitation spectrum reorganizes via a level crossing. Evaluating these mean-field predictions using the parameters of Fig.~\ref{fig1}, we obtain $U_c/J=-0.0625$ for $P=0$ [panel (c)] and $U_c/J=-0.5$ for $P\neq 0$ [panel (d)], in very good agreement with the values obtained numerically by exact diagonalization and associated to the local maxima of $S_\text{ent}$.

\begin{table}[t]
\caption{Stationary points of the classical energy $\mathcal{E}(\phi,z)$}
\label{tab1}
\renewcommand{\arraystretch}{1.14}
\begin{ruledtabular}
\begin{tabular}{cccc}
Region & Minima & Saddles & Maxima \\
\hline
A 
& $(0,\pm z_s)$ 
& $(0,0)$ 
& $(\pm\pi,0)$ \\

B 
& $(0,0)$ 
& --- 
& $(\pm\pi,0)$ \\

C 
& $(0,0)$ 
& $(\pm\pi,0)$ 
& $(\pm\pi,\pm z_s)$ \\

$\overline{\mathrm{A}}$ 
& $(0,\pm z_s)$ 
& $(\pm\phi_s,0)$ 
& $(0,0)$, $(\pm \pi,0)$ \\

$\overline{\mathrm{B}_1}$ 
& $(\pm\phi_s,0)$ 
& $(0,\pm z_s)$ 
& $(0,0)$, $(\pm\pi,0)$ \\

$\overline{\mathrm{B}_2}$ 
& $(\pm\phi_s,0)$ 
& $(0,0)$ 
& $(\pm\pi,0)$ \\

$\overline{\mathrm{C}}$ 
& $(\pm\phi_s,0)$ 
& $(0,0)$, $(\pm \pi,0)$ 
& $(\pm\pi,\pm z_s)$ \\
\end{tabular}
\end{ruledtabular}
\end{table}

\begin{figure}
    \centering
    \includegraphics[width=\linewidth]{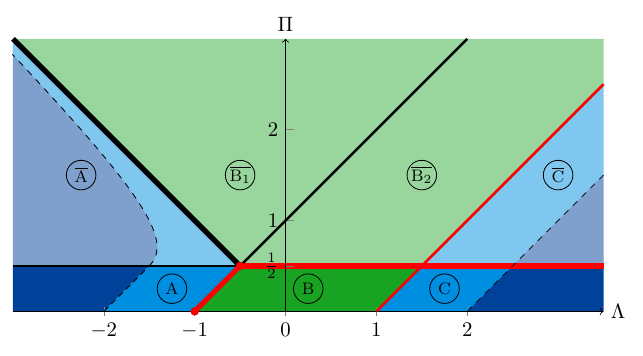}
    \caption{\textbf{Mean-field phase diagram (see also Fig.~\ref{fig9} for an extended version).} Solid lines delineate regions characterized by different stationary points of $\mathcal E(\phi, z)$. Thick solid lines indicate boundaries across which the minima change, either continuously (red lines) or discontinuously (black lines), while thin solid lines indicate boundaries across which the maxima change, either continuously (red lines) or discontinuously (black lines). Different colors denote the three dynamical regimes described in Sec.~\ref{sec:MFdyn}: Josephson regime (green), phase-locked MQST (light blue), and running-phase MQST (dark blue). Above the line $\Pi=\frac{1}{2}$ the same dynamical regimes are present with qualitative differences.}
    \label{fig2}
\end{figure}

\begin{figure*}
    \centering
    \includegraphics[width=\linewidth]{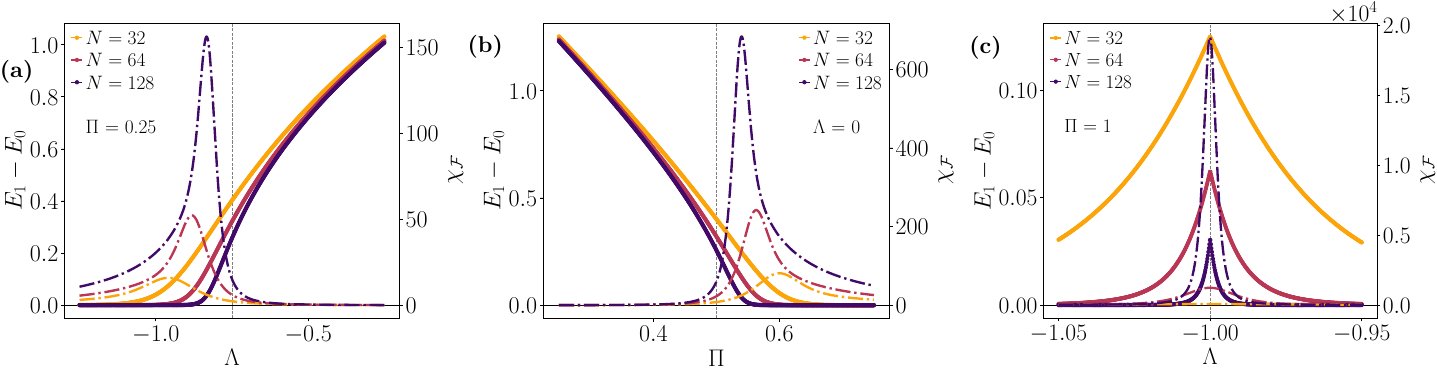}
    \caption{\textbf{Equilibrium quantum phase transitions}. Thick solid lines (circular markers) denote the energy gap $E_1-E_0$, while thin dashed-dotted lines indicate the ground-state fidelity susceptibility $\chi_{\mathcal F}$ for several values of $N$. The dashed vertical line at the center of each panel represents the mean-field critical value of the driving parameter. (a) $\Lambda$-driven continuous QPT to NOON-like states in the weak pair tunneling regime ($\Pi = 0.25$), where $\Lambda_c=-0.75$. (b) $\Pi$-driven continuous QPT to phase-NOON-like states for $\Lambda=0$, where $\Pi_c=0.5$. (c) $\Lambda$-driven first-order QPT to from phase-NOON-like to NOON-like states in the strong pair tunneling regime ($\Pi=1.0$), where $\Lambda_c=-1.0$. Parameter $J=1.0$.}
    \label{fig3}
\end{figure*}

Finally, for $\Lambda > -\frac{1}{2}$, by crossing the line
\begin{equation}
    \Pi_c=\frac{1}{2}
\end{equation}
the minimum of $\mathcal E(\phi,z)$ changes continuously from $(0,0)$ to $(\pm \phi_s, 0)$, which signals the pair-tunneling-induced continuous QPT. All this information is summarized in the zero-temperature mean-field phase diagram represented in Fig.~\ref{fig2}. The values of the stationary points of $\mathcal E(\phi, z)$ in each region, which will also be relevant for the mean-field dynamics (Sec.~\ref{sec:MFdyn}), are reported in Table \ref{tab1}. An extended version of the phase diagram, including illustrative contour plots of $\mathcal{E}(\phi,z)$ for each region, is provided in Fig.~\ref{fig9} (Appendix~\ref{app:fig}).

The above mean-field description can be directly connected to the properties of the quantum ground state \cite{Polls, Zhu, Liu}. At any finite $N$, the exact many-body ground state is nondegenerate, real in the Fock basis, and fully respects the symmetries of the Hamiltonian. For $\Lambda < \Lambda_c$, in the large-$N$ limit the ground state $\ket{E_0}$ and the first excited state $\ket{E_1}$ are given, respectively, by symmetric and antisymmetric superpositions of wave packets localized around the two imbalanced semiclassical ground states $(0, \pm z_s)$. These wave packets are well described by the $SU(2)$ atomic coherent states---namely, minimal-uncertainty product states with well-defined relative phase $\phi_0$ and population imbalance $z_0$ \cite{Arecchi, Leggett}---
\begin{equation}\label{acs}
    \ket{(\phi_0, z_0)}\!=\!\frac{1}{\sqrt{N!}}\!\left(\!\sqrt{\frac{1\!+\!z_0}{2}}\hat a_L^\dag \!+\! \sqrt{\frac{1\!-\!z_0}{2}}e^{\iunit \phi_0} \hat a_R^\dag \!\right)^N\!\ket{0,0},
\end{equation}
with $\phi_0=0$ and $z_0=\pm z_s$ \footnote{The atomic coherent state is represented by the $N$-body wavefunction $\Psi(\{x_i\}) = \prod_{i=1}^N [\sqrt{\frac{1+z_0}{2}}\Phi_L(x_i)+\sqrt{\frac{1-z_0}{2}}e^{\iunit \phi_0} \Phi_R(x_i)]$. The values of $z_0$ and $\phi_0$ can be measured in a time-of-flight expansion. If at $t=0$ the trap is released, maintaining only the confinement in the transverse directions, the condensate wavefunctions evolve freely, and at time $t>0$ they are given by $\Phi_{L(R)}(x,t) \propto \tilde \Phi(\frac{mx}{\hbar t})e^{\iunit mx^2/2\hbar t}e^{\pm \iunit max/2\hbar t}$, where $a$ is the separation between the two wells and $\tilde \Phi(k)$ denotes the Fourier transform. The single-particle density $\rho_1(x, t) = \int dx_2\cdots dx_N\,|\Psi(x,x_2,\dots,x_N, t)|^2 \propto |\tilde \Phi(\frac{mx}{\hbar t})|^2 [1 + \sqrt{1-z_0^2} \cos(\phi_0-\frac{max}{\hbar t})]$ thus shows interference fringes whose contrast and phase depend on $z_0$ and $\phi_0$.}. In the strongly attractive regime, $\Lambda \ll \Lambda_c$, these wave packets become sharply localized around the Fock states $\ket{N,0}$ and $\ket{0,N}$, so that $\ket{E_0}$ and $\ket{E_1}$ approach the symmetric and antisymmetric NOON states $\ket{N,0} \pm \ket{0,N}$. The degeneracy of the semiclassical ground states corresponds to the fact that the system in the quantum ground state is a fragmented condensate \cite{Leggett, Penrose, Zhu, Baym, Fischer1, Fischer2}, characterized by two macroscopic eigenvalues of the one-body density matrix
\begin{equation}\label{one-body}
    \hat\rho^{(1)} = \begin{pmatrix}
        n_L & \frac{N\alpha}{2} \\[1ex] \frac{N\alpha^*}{2} & n_R
    \end{pmatrix},
\end{equation}
where $n_{L(R)} \equiv \langle \hat n_{L(R)}\rangle$ and $\alpha \equiv \frac{2}{N}\langle \hat a_L^\dag \hat a_R\rangle$, with $\langle \cdots\rangle$ denoting in this case the expectation value on $\ket{E_0}$. In the thermodynamic limit, the energy splitting between the symmetric and antisymmetric superpositions vanishes exponentially with $N$. An arbitrarily weak perturbation selects one of the semiclassical minima, leading to spontaneous breaking of the left-right symmetry.

A similar situation occurs for $\Lambda > \Lambda_c$ in the presence of strong pair tunneling, $\Pi > \frac{1}{2}$. In this regime, in the large-$N$ limit the two lowest-energy eigenstates are given by symmetric and antisymmetric superpositions of wave packets localized around the phase-imbalanced semiclassical ground states $(\pm \phi_s, 0)$, namely $\ket{(\phi_s,0)} \pm\ket{(-\phi_s,0)}$. Deep in pair tunneling regime, $\Pi \gg \frac{1}{2}$, these approach $\ket{(\frac{\pi}{2},0)} \pm \ket{(-\frac{\pi}{2},0)}$, which may be interpreted as symmetric and antisymmetric phase-NOON states, again corresponding to a fragmented condensate \cite{Zhu}. In the thermodynamic limit, the degeneracy of the two eigenstates leads to spontaneous symmetry breaking, in this case associated with the selection of one of the two time-reversal-related semiclassical ground states.

The mean-field analysis is corroborated by Fig.~\ref{fig3}, which presents exact numerical results for the behavior of the gap $E_1-E_0$ and the ground-state fidelity susceptibility $\chi_{\mathcal F}$, defined as \cite{Gu_rev}
    \begin{equation}
    \label{fidelity}
    \chi_{\mathcal F}(g) = \lim_{\delta g \to 0} \frac{-2 \ln |\braket{E_0(g)}{E_0(g+\delta g)}|}{\delta g^2},
\end{equation}
where $g$ denotes the control parameter driving the QPT. Since the parameter space is two-dimensional, described by $(\Lambda, \Pi)$ or equivalently by $(U, P)$, any trajectory connecting two quantum phases, and thus crossing a line of QPTs, can be linearized in the vicinity of the transition point as a straight line, so that $g$ can be expressed as a linear combination of $U$ and $P$. The precise relation between these parameters depends on the specific choice of the trajectory and is immaterial for the identification of the transitions themselves. The fidelity susceptibility quantifies the sensitivity of the ground state to an infinitesimal variation of the control parameter and thus provides a faithful probe of quantum critical behavior. For continuous QPTs, the excitation gap closes in the thermodynamic limit; at finite $N$, although the gap does not vanish exactly, it is expected to become extremely small in the symmetry-broken phase, reflecting the emergence of quasi-degenerate ground states associated with spontaneous symmetry breaking. Correspondingly, the fidelity susceptibility develops a pronounced peak at a pseudo-critical point that approaches the mean-field critical value as $N$ increases, with a peak height that grows algebraically and a width that narrows with increasing $N$ \cite{Zanardi, You, Zanardi2, Gu, Rigol, Liu, Gu_rev}. 

These features are clearly observed in Figs.~\ref{fig3}(a)-(b), which display the $\Lambda$-driven and $\Pi$-driven continuous transitions to NOON-like and phase-NOON-like states, respectively, and thus confirm the mean-field predictions for both the location and the continuous nature of the transitions. By contrast, Fig.~\ref{fig3}(c), obtained for $\Pi > \frac{1}{2}$, exhibits qualitatively different behavior: the gap displays a cusp, while the fidelity susceptibility shows a very narrow and rapidly growing peak. This behavior is characteristic of a first-order QPT, where the gap closing in the thermodynamic limit originates from a level crossing between two macroscopically distinct ground states rather than from the softening of a collective excitation mode. At finite $N$ the transition is replaced by a sharp avoided crossing \cite{Campostrini}, with a minimum gap that decreases approximately linearly with increasing $N$. The fact that the gap vanishes asymptotically away from the transition, both for $\Lambda < \Lambda_c$ and for $\Lambda > \Lambda_c$, reflects the first-order nature of the transition between two distinct symmetry-broken phases, characterized by broken left-right symmetry for $\Lambda < \Lambda_c$ and broken time-reversal symmetry for $\Lambda > \Lambda_c$.

\begin{figure*}
    \centering
    \includegraphics[width=\linewidth]{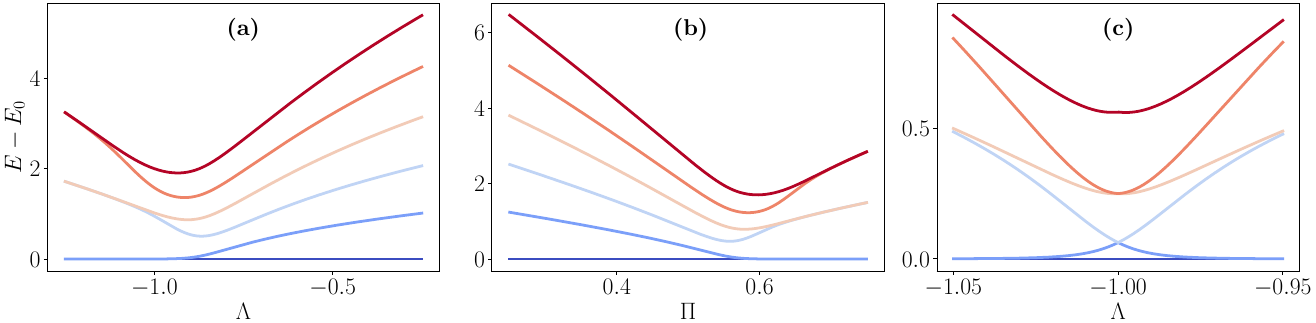}
    \caption{\textbf{Low-energy spectrum}. Energy of the first five excited states with respect to the energy of the ground state, for $N=64$ and the same parameters of Fig.~\ref{fig3}. (a) $\Pi=0.25$. (b) $\Lambda=0$. (c) $\Pi=1.0$. Parameter $J=1.0$.}
    \label{fig4}
\end{figure*}

This analysis is complemented by Fig.~\ref{fig4}, which displays the first few energy levels over the same parameter ranges considered in Fig.~\ref{fig3}. We observe that, deeper within both symmetry-broken phases, the merging of energy levels also occurs for higher consecutive pairs, e.g. $E_3 \simeq E_2$, $E_5 \simeq E_4$, etc. \cite{Cirac, Vianello}. The avoided crossings between such pairs of eigenvalues at the first-order pseudo-critical point are also clearly visible in Fig.~\ref{fig4}(c).

\section{Dynamics}\label{sec:dyn}

We now turn to the effects of pair tunneling on the dynamics. We initialize the system in a pure state $\ket{\Psi_0}= \sum_{i=0}^N \psi_i \ket{i, N-i}$ and consider its unitary time evolution $\ket{\Psi(t)} = e^{-\iunit \hat H t/\hbar}\ket{\Psi_0}=\sum_{i=0}^N \psi_{i}(t)\ket{i,N-i}$. A naturally interesting class of initial states is provided by the atomic coherent states $\ket{(\phi_0, z_0)}$ \eqref{acs}, for which $\psi_i = \sqrt{\binom{N}{i}\frac{1}{2^N}}(1+z_0)^{i/2}(1-z_0)^{(N-i)/2}e^{\iunit(N-i)\phi_0}$. In fact, these states are the ones most commonly realized in experiments, where the large number of atoms typically ensures dynamics close to the mean-field description. By considering atomic coherent states, we thus maximize the likelihood of identifying phenomena that persist at experimentally relevant particle numbers, rather than effects that are purely few-body in nature and potentially difficult to observe. Moreover, the choice of $\ket{(\phi_0, z_0)}$ as initial state allows for a direct comparison between the ensuing quantum dynamics and the mean-field dynamics of Eq.~\eqref{semiclas} with initial condition $(\phi_0, z_0)$ \cite{Meystre, Vianello2}.

\subsection{Mean-field dynamics: Josephson oscillations and Macroscopic Quantum Self-Trapping}\label{sec:MFdyn}

The mean-field dynamics can be inferred from the phase-space structure of the stationary points of the energy $\mathcal E(\phi,z)$. While the location (and possible degeneracy) of the local minima determines the ground-state properties, the different dynamical regimes are governed by the overall arrangement of local minima, saddle points, and local maxima of $\mathcal E(\phi,z)$, see Figs.~\ref{fig2} and \ref{fig9}. Let us recall the salient features of the mean-field dynamics when pair tunneling is absent: $\Pi=0$ \cite{Mele-Messeguer, Raghavan}. Since $\mathcal E(\phi,z)$ is symmetric under the transformation $\phi \to -\phi$, we may restrict attention to the strip $[0,\pi]\times[-1,1]$. We consider first the case of repulsive interactions. For $0 \le \Lambda <1$ all trajectories are closed, representing Josephson oscillations around the minimum $(0,0)$ or the maximum $(\pi,0$). At $\Lambda=1$, the point $(\pi,0)$ undergoes a pitchfork bifurcation and becomes a saddle, while new maxima appear at $(\pi, \pm z_s)$. Hence the dynamics consists of Josephson oscillations around the minimum, with $\langle z(t)\rangle_t=0$, or closed orbits around the maxima with $\langle z(t)\rangle_t\neq 0$; the latter is the phase-locked ($\pi$-phase) macroscopic quantum self-trapping (MQST). Finally, for $\Lambda >2$, separatrices connecting the saddle points $(\pm\pi, 0)$ emerge, and a new region opens up between the separatrices and the libration regions of $(\pi, \pm z_s)$, where the system can explore open trajectories with $\langle z(t)\rangle_t \neq 0$, namely running-phase MQST. Increasing further the energy, trajectories may enter the small regions surrounding $(\pi, \pm z_s)$, giving rise again to $\pi$-phase MQST. The minimal requirement for which some initial conditions produce MQST is therefore $\Lambda > 1$. More generally, given the initial conditions $(\phi_0, z_0)$, MQST occurs when the energy of the trajectory exceeds the saddle energy, i.e. $\mathcal E(\phi_0, z_0)>\mathcal E(\pi, 0)$. This yields $\Lambda> 2(\sqrt{1-z_0^2}\cos\phi_0+1)/z_0^2$.

Since the transformation $\phi \to \pi-\phi$, $\Lambda\to-\Lambda$ maps $\mathcal E \to -\mathcal E$, the phase-space portraits for attractive and repulsive interactions are identical up to a phase shift of $\pi$ combined with time reversal. Therefore the same dynamical regimes occur for attractive interactions; in particular, the condition for MQST is $\mathcal E(\phi_0, z_0)<\mathcal E(0,0)$, which yields $\Lambda < 2(\sqrt{1-z_0^2} \cos\phi_0-1)/z_0^2$.

Let us now examine the effects of pair tunneling. For weak pair tunneling, $\Pi \le \frac{1}{2}$, the picture discussed above remains qualitatively unchanged, as the same fixed points and dynamical regimes are present; see Table \ref{tab1} and Fig.~\ref{fig2}. Phase-locked MQST can occur for $\Lambda > 1+\Pi$ or $\Lambda <\Lambda_c=-1+\Pi$ (light blue regions in Fig.~\ref{fig2}), while running-phase MQST can occur for $\Lambda > 2+\Pi$ or $\Lambda < -2+\Pi$ (dark blue regions in Fig.~\ref{fig2}), provided the initial conditions satisfy
\begin{equation*}
    \Lambda > \frac{\Pi [1\!-\!(1\!-\!z_0^2)\cos 2\phi_0] \!+\! 2(\sqrt{1\!-\!z_0^2}\cos\phi_0 \!+\! 1)}{z_0^2}
\end{equation*}
or
\begin{align}\label{conmqst}
    &\Lambda < \frac{\Pi [1\!-\!(1\!-\!z_0^2)\cos 2\phi_0] \!+\! 2(\sqrt{1\!-\!z_0^2}\cos\phi_0 \!-\! 1)}{z_0^2},\nonumber\\
    &(0 \le \Pi \le \textstyle \frac{1}{2}),
\end{align}
for repulsive and attractive interactions, respectively. Therefore MQST is not the only dynamical regime accessible in these regions of parameter space. If the requirements on the initial conditions are not satisfied, Josephson dynamics occurs instead.

The situation changes for $\Pi > \frac{1}{2}$, where new phase-imbalanced fixed points appear at $(\pm \phi_s, 0)$, that are saddles for $\Lambda < -\Pi$ and minima for $\Lambda > -\Pi$. In the former case, Josephson oscillations can occur around the maximum at $(0,0)$ even for negative interaction strengths. In the latter case, oscillations may take place around each of the two minima at $(\pm \phi_s, 0)$ individually---corresponding to phase-imbalanced Josephson oscillations with $\langle \phi(t)\rangle_t \neq \{0, \pi\}$---or around both simultaneously. In general, phase-locked MQST can occur for ${\Lambda > 1+\Pi}$ or $\Lambda <  \Lambda_c = -\Pi$ (light blue regions in Fig.~\ref{fig2}), while running-phase MQST can occur for ${\Lambda >2+\Pi}$ or $\Lambda < -\frac{1}{2\Pi}-\Pi$ (dark blue regions in Fig.~\ref{fig2}), provided that $\mathcal E(\phi_0, z_0) > \mathcal E(\pi,0)$ or $\mathcal E(\phi_0, z_0) < \mathcal E(\phi_s,0)$, namely
\begin{equation*}
    \Lambda > \frac{\Pi [1\!-\!(1\!-\!z_0^2)\cos 2\phi_0] \!+\! 2(\sqrt{1\!-\!z_0^2}\cos\phi_0 \!+\! 1)}{z_0^2}
\end{equation*}
or
\begin{align}
    &\Lambda < \frac{-\Pi [1\!+\!(1\!-\!z_0^2)\cos 2\phi_0] \!+\! 2(\sqrt{1\!-\!z_0^2}\cos\phi_0 \!-\! \frac{1}{4\Pi})}{z_0^2} ,\nonumber\\
    &(\Pi > \textstyle \frac{1}{2}),
\end{align}
for repulsive and attractive interactions, respectively. Notice that for repulsive interactions, the condition is the same as in Eq.~\eqref{conmqst}. 

\subsection{Comparison with the quantum dynamics}

The mean-field dynamics described in Sec.~\ref{sec:MFdyn} is expected to accurately reproduce the quantum dynamics for large $N$ at fixed $\Lambda$ and $\Pi$, becoming exact in the thermodynamic limit $N \to \infty$. Formally, this follows from the coherent-state path-integral representation of transition amplitudes, $\int \mathcal D\phi \mathcal Dz\,e^{\iunit N s(\phi,z)/\hbar}$, in which the action is naturally multiplied by $N$, see Eq.~\eqref{action}. In the limit $N\to\infty$ the path integral is therefore dominated by stationary phase trajectories, which satisfy the equations of motion \eqref{semiclas} \cite{Vianello2}. Finite-$N$ deviations from the mean field, arising from genuine few-body quantum effects, can be studied by directly comparing the mean-field dynamics initialized at $(\phi_0, z_0)$ with the full quantum evolution of the corresponding atomic coherent state $\ket{(\phi_0, z_0)}$, that is $\ket{\Psi(t)}=e^{-\iunit\hat Ht/\hbar}\ket{(\phi_0, z_0)}$. The expectation value of the population imbalance is given by
\begin{equation}
    z(t) = \frac{\braop{\Psi(t)}{\hat n_L-\hat n_R}{\Psi(t)}}{N}.
\end{equation}
The relative phase is extracted from the one-body density matrix $\rho^{(1)}_{ij}(t) = \braop{\Psi(t)}{\hat a_i^\dag \hat a_j}{\Psi(t)}$, $i,j\in \{L, R\}$ [Eq.~\eqref{one-body}]. For nonzero $\alpha(t) = \frac{2}{N}\braop{\Psi(t)}{\hat a_L^\dag \hat a_R}{\Psi(t)}$, the eigenvalues $\varrho_\pm(t)$ of $\hat\rho^{(1)}$ satisfy
\begin{equation}
    \frac{\varrho_\pm(t)}{N} = \frac{1 \pm \sqrt{z(t)^2+|\alpha(t)|^2}}{2}
\end{equation}
and the corresponding eigenvectors are
\begin{equation}
    \ket{\chi_\pm(t)} \propto \begin{pmatrix} \alpha(t) \\[1ex] -z(t) \pm \sqrt{z(t)^2+|\alpha(t)|^2} \end{pmatrix},
\end{equation}
up to normalization. The largest-eigenvalue eigenvector of $\hat\rho^{(1)}$ is the complex conjugate of the macroscopically occupied single-particle state \cite{Leggett, Penrose}, therefore the relative phase between the condensates in the two wells is the argument of $\alpha(t)$, namely
\begin{equation}
    \phi(t) = \text{arg}\,\braop{\Psi(t)}{\hat a_L^\dag \hat a_R}{\Psi(t)}.
\end{equation}

\begin{figure}
    \centering
    \includegraphics[width=\linewidth]{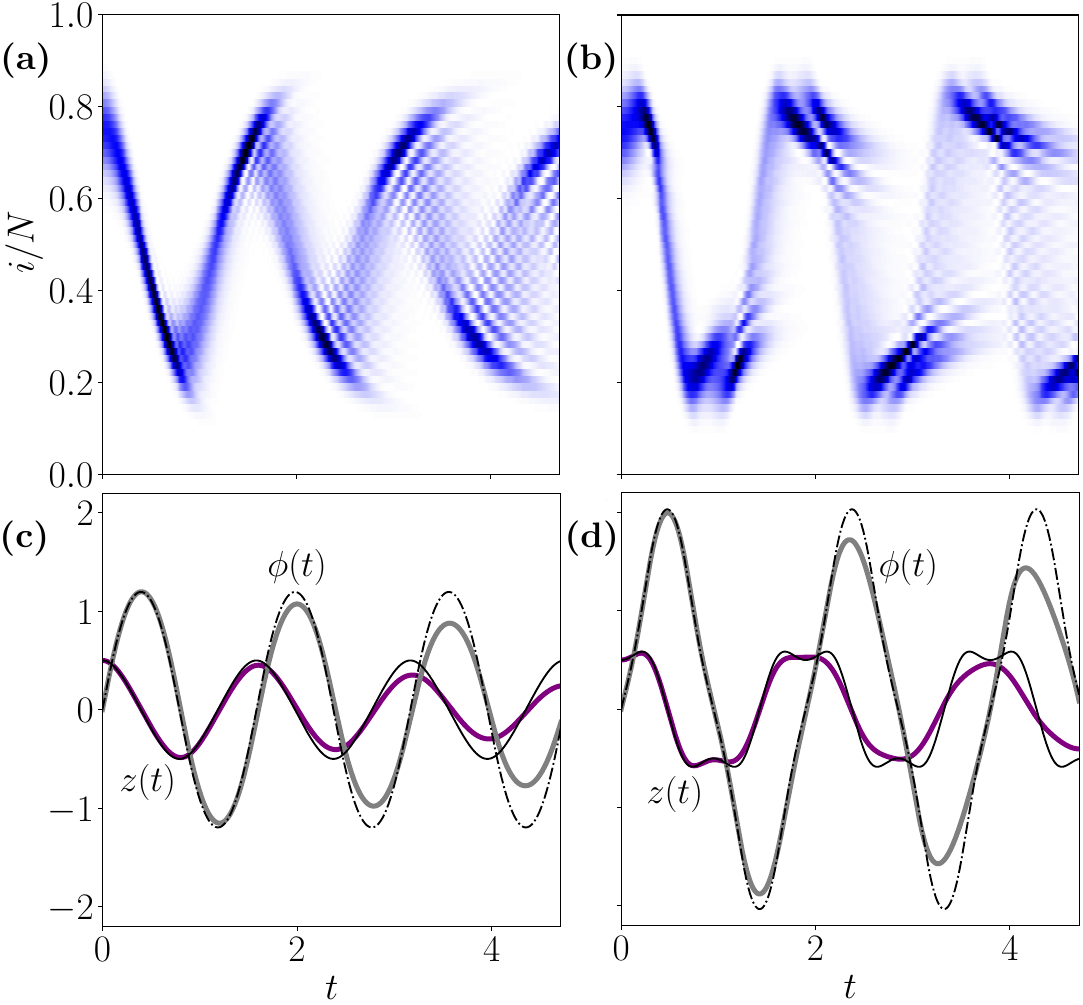}
    \caption{\textbf{Josephson oscillations.} Time evolution of the probabilities $|\psi_i(t)|^2$ [panels (a)-(b)], and of the population imbalance $z(t)$ and the relative phase $\phi(t)$ [panels (c)-(d)] for the system initialized in an atomic coherent state with $z_0=0.5$ and $\phi_0=0$. Darker blue corresponds to higher probability. Exact quantum results (bold solid lines) are compared with the corresponding mean-field trajectories (thin lines). (a) and (c): $\Lambda=4.0$, $\Pi=0$. (b) and (d): $\Lambda=4.0$, $\Pi=1.2$. Parameters are $N=64$ and $J=1.0$. Time is in units of $\hbar/J$.}
    \label{fig5}
\end{figure}

\begin{figure}
    \centering
    \includegraphics[width=\linewidth]{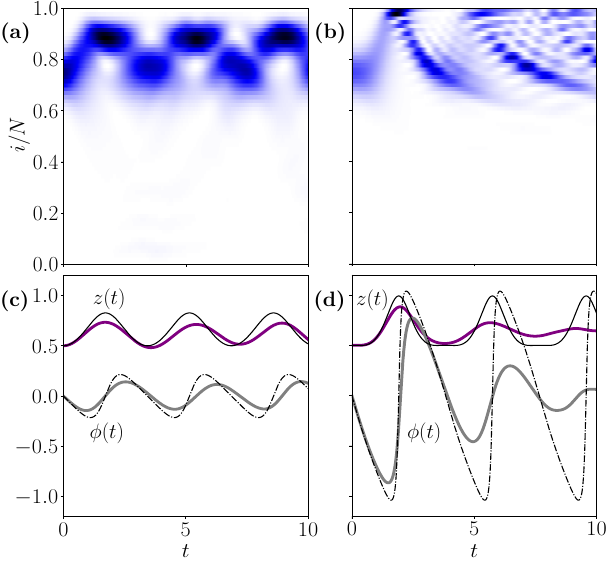}
    \caption{\textbf{Phase-locked MQST.} Time evolution of the probabilities $|\psi_i(t)|^2$ [panels (a)-(b)], and of the population imbalance $z(t)$ and the relative phase $\phi(t)$ [panels (c)-(d)] for a system initialized in the atomic coherent state with $z_0=0.5$ and $\phi_0=0$. Darker blue corresponds to higher probability. Exact quantum results (bold solid lines) are compared with the corresponding mean-field trajectories (thin lines). (a) and (c): $\Lambda=-1.4$, $\Pi=0$. (b) and (d): $\Lambda=-1.4$, $\Pi=0.6$. Parameters are $N=64$ and $J=1.0$. Time is in units of $\hbar/J$.}
    \label{fig6}
\end{figure}

\begin{figure}
    \centering
    \includegraphics[width=\linewidth]{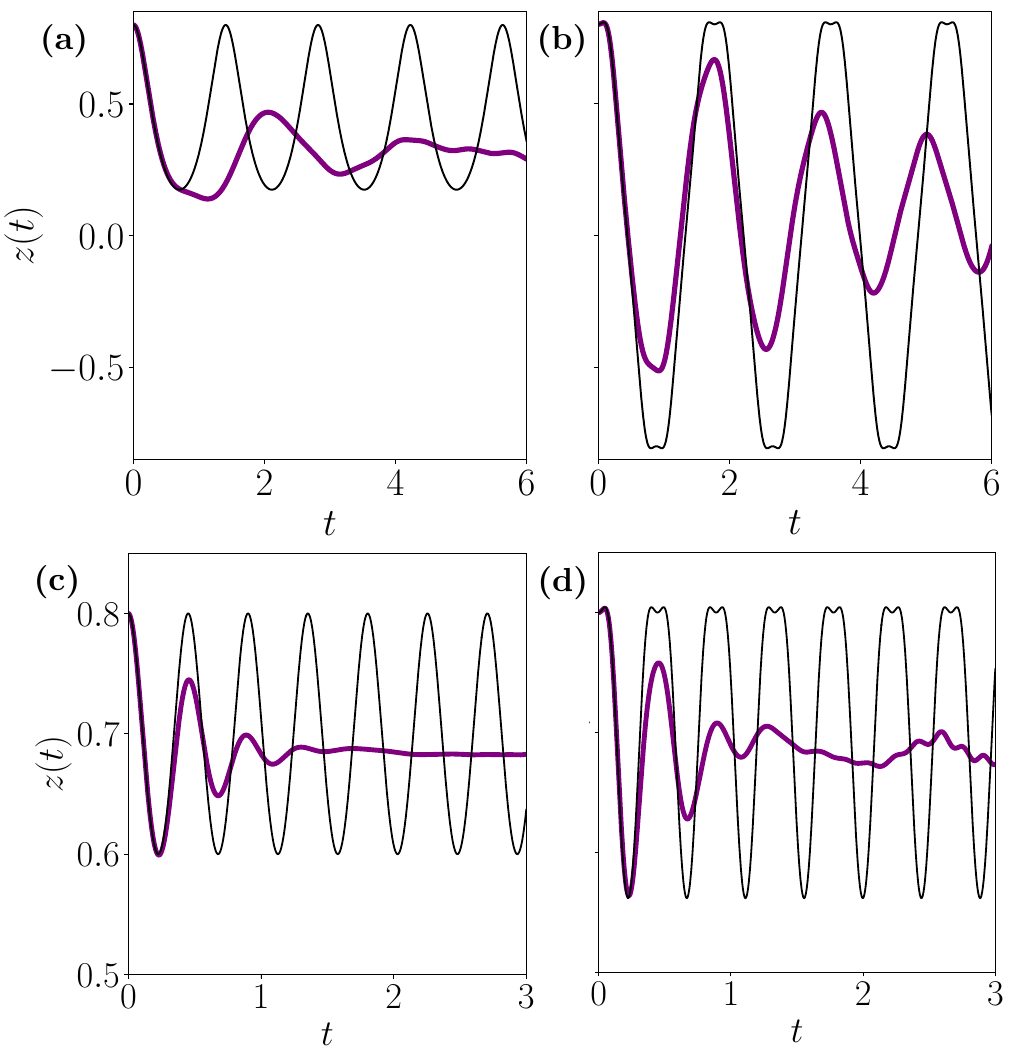}
    \caption{\textbf{Running-phase MQST.} Time evolution of the population imbalance $z(t)$ for a system initialized in an atomic coherent state with $z_0=0.8$ and $\phi_0=0$ (bold solid lines), compared with the corresponding mean-field dynamics (thin lines). (a) $\Lambda=5.2$, $\Pi=0$. (b) $\Lambda=5.2$, $\Pi=1.2$. (c) $\Lambda=10.0$, $\Pi=0$. (d) $\Lambda=10.0$, $\Pi=1.2$. Parameters are $N=64$ and $J=1.0$. Time is in units of $\hbar/J$.}
    \label{fig7}
\end{figure}

In Fig.~\ref{fig5} we report an example of Josephson oscillations for $\Lambda >0$, comparing dynamics in the absence and the presence of pair tunneling, as well as quantum and mean-field descriptions. Alongside the time evolution of the macroscopic observables $z(t)$ and $\phi(t)$, we show the underlying microscopic dynamics through the evolution of the Fock-state occupation probabilities $|\psi_i(t)|^2$. Here, as in Figs.~\ref{fig6} and \ref{fig7}, the interaction parameters are chosen to enhance the distinction between the mean-field and exact quantum evolutions and to make the relevant dynamical features as clear and legible as possible. For the chosen interaction parameters and initial conditions, the inclusion of pair tunneling leads to Josephson oscillations around both minima at $(\pm\phi_s,0)$, resulting in a markedly non-sinusoidal evolution of $z(t)$ characterized by a doubly peaked profile, which is also reflected in the structure of the Fock-state probabilities. The mean-field trajectories are in good agreement with the exact quantum results at short times, accurately reproducing the initial oscillation frequency and overall behavior. At longer times, few-body effects---such as amplitude modulations and increasingly anharmonic features---become evident \cite{Sakmann, Vianello2}. These effects are accompanied by a progressive spreading of the wave packet in the Fock basis, as visible in Fig.~\ref{fig5}(a)-(b), signaling a reduction of its coherence. In general, for a fixed particle number $N$, the agreement between exact and mean-field descriptions improves and persists over longer times for weaker interactions. In the noninteracting limit $\Lambda=\Pi=0$, as far as the observables $\phi(t)$ and $z(t)$ are concerned the quantum dynamics of an atomic coherent state follows exactly the mean field dynamics at any $N$ \cite{Vianello2}. 

Similarly, Fig.~\ref{fig6} illustrates an instance of phase-locked MQST for $\Lambda <0$, where the mean-field dynamics corresponds to closed trajectories around the fixed point $(0, z_s)$. In this regime, the most striking effect of pair tunneling appears at the microscopic level: the additional nonlinearity substantially reshapes the Fock-state probabilities, leading to a qualitatively different distribution compared with the case without pair tunneling, despite the persistence of self-trapped behavior.

In Fig.~\ref{fig7} we present an example of the third dynamical regime, namely running-phase MQST, for $\Lambda > 0$. For the chosen initial conditions, the mean-field criterion for the occurrence of MQST reads $\Lambda-\Pi >5.0$. In panels (a) and (b) we fix $\Lambda=5.2$ and compare the cases $\Pi=0$ and $\Pi=1.2$. The inclusion of pair tunneling lowers the energy of the mean-field trajectories below that of the separatrix, resulting in the expected destruction of MQST and its conversion into large-amplitude Josephson oscillations around both minima at $(\pm\phi_s,0)$. This prediction is confirmed by the exact numerical results; however, for these moderately large values of $\Lambda$, the quantum dynamics rapidly departs from the mean-field behavior. In panels (c) and (d), we consider instead a parameter regime in which the inclusion of pair tunneling is not expected to destroy self-trapping. In this case as well, the quantum results are in qualitative agreement with this mean-field prediction, although the strong interactions cause the quantum dynamics to become incoherent at short times, i.e. smaller than one semiclassical period.

\subsection{Dynamical quantum phase transitions}

We complete our characterization of the effects of pair tunneling by considering dynamical quantum phase transitions. These extend the notion of criticality to the time domain \cite{Heyl1, Zvyagin, Heyl2}, where the central object is the Loschmidt amplitude
\begin{equation}\label{Losch_amp}
    \mathcal Z(t) = \braop{\Psi_0}{e^{-\iunit \hat Ht/\hbar}}{\Psi_0},
\end{equation}
giving the overlap between the initial state $\ket{\Psi_0}$ and its time-evolved counterpart at time $t$. The associated probability, $\mathcal L(t) = |\mathcal Z(t)|^2$, is known as the Loschmidt echo or fidelity. The formal similarity of the Loschmidt amplitude to the partition function $Z(\beta)$ of an equilibrium system at inverse temperature $\beta$ motivates the definition of DQPTs paralleling that of thermal phase transitions. In equilibrium, criticality emerges in the thermodynamic limit as a nonanalyticity of the free energy density $f(\beta) = -\frac{1}{\beta N}\ln Z(\beta)$ at a critical temperature $\beta=\beta_c$. Correspondingly, a DQPT is defined in the thermodynamic limit as a nonanalyticity (typically a cusp) of the return rate
\begin{equation}\label{lambda}
    \lambda(t)=-\frac{1}{N} \ln |\mathcal Z(t)|^2
\end{equation}
at a critical time $t=t_c$.

Inserting the eigenbasis $\ket{E_n}$ in Eq.~\eqref{Losch_amp} and introducing the imaginary time ${\tau = \iunit t}$, the Loschmidt amplitude can be written as $\mathcal Z(\tau)=\sum_{n=0}^N |\braket{E_n}{\Psi_0}|^2 e^{-E_n\tau/\hbar}$. For finite $N$, this is a finite sum of analytic functions and is therefore itself analytic. As a consequence, it can be factorized as \cite{Heyl2}
\begin{equation}\label{Losch_zeros}
    \mathcal Z(\tau)=e^{c\tau}\prod_j \left(1-\frac{\tau}{\tau_j}\right),
\end{equation}
where $c$ is a constant and $\tau_j$ are the complex zeros of the Loschmidt amplitude (Loschmidt zeros). These are the nonequilibrium analogue of Fisher or Lee-Yang zeros of thermal partition functions \cite{Fisher, Yang1, Yang2} and determine the nonanalytic properties of the return rate, as
\begin{equation}
    \lambda(\tau) = -\frac{2}{N}\,\text{Re} \biggl[c\tau+\sum_j \ln \left(1-\frac{\tau}{\tau_j}\right)\biggr].
\end{equation}
For finite $N$, the Loschmidt zeros are isolated points in the complex plane. As $N$ increases, they become increasingly dense and, in the thermodynamic limit, coalesce into continuous lines or regions. Intersections of these structures with the imaginary $\tau$ axis determine the real critical times at which the return rate becomes nonanalytic and DQPTs occur. The occurrence of a DQPT (and the associated critical time) is therefore tied to whether a state orthogonal to the initial state---i.e., maximally separated in the Hilbert space---is dynamically accessible under time evolution, and if so, at which time.

Since its introduction in the context of the transverse-field Ising model by Heyl \emph{et al.} \cite{Heyl1}, the study of DQPTs has extended to a variety of many-body systems \cite{Weidinger, Halimeh}, including topological systems \cite{Vajna2, Schmitt, Lahiri}, systems with long-range interactions \cite{Silva, Syed}, superfluids and superconductors \cite{Abdi, Domanski}. The phenomenon has also been observed experimentally in an interacting trapped-ion platform \cite{Jurevic}. While the original work of Hely \emph{et al.} suggested a close connection between DQPTs and equilibrium QPTs, this correspondence is not universal \cite{Vajna}: a DQPT may occur without crossing an equilibrium critical point and, conversely, a quench may cross an equilibrium critical point without generating a DQPT. As we shall show, our system falls into this more general scenario, exhibiting dynamical critical behavior that is not tied to quenches across equilibrium criticality, such as tuning $\Lambda$ from $\Lambda < \Lambda_c$ to $\Lambda > \Lambda_c$. In this case, DQPTs reflect genuinely dynamical features of the real-time evolution, with no direct counterpart in equilibrium critical behavior.

We address the identification of DQPTs by means of two complementary numerical approaches. First, we compute the Loschmidt echo and the associated return rate directly from the full energy spectrum obtained via exact diagonalization, and analyze their scaling with $N$. In particular, we search for finite-$N$ precursors of nonanalytic cusps in the return rate when the Loschmidt echo approaches zero. To substantiate the signatures inferred from this direct inspection, we also determine the complex Loschmidt zeros in the vicinity of the imaginary $\tau$ axis using the method of Loschmidt cumulants \cite{Peotta, Zhao}, outlined in Appendix \ref{appA}. 

\begin{figure*}
    \centering
    \includegraphics[width=\linewidth]{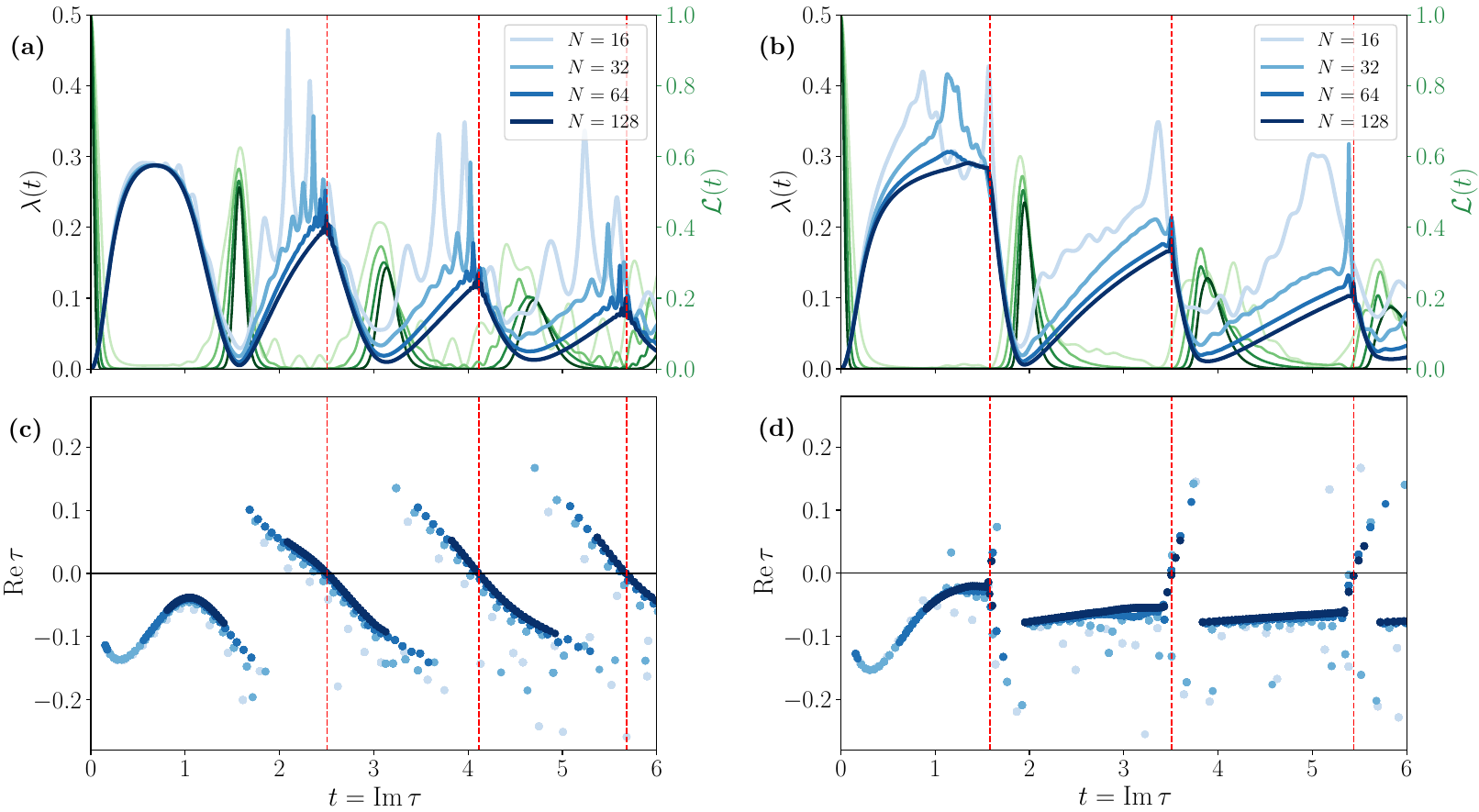}
    \caption{\textbf{Dynamical quantum phase transitions.} Time evolution of the Loschmidt echo $\mathcal L(t)$ (green lines) and the return rate $\lambda(t)$ (blue lines) [panels (a)-(b)], and complex zeros of the Loschmidt amplitude in the vicinity of the imaginary $\tau$ axis [panels (c)-(d)], for a system initialized in an atomic coherent state with $\phi_0=0$ and $z_0=0.5$ and several values of $N$. Red dashed lines identify the critical times. (a) and (c): $\Lambda = 4.0$, $\Pi=0$. (b) and (d): $\Lambda=4.0$ and $\Pi = 0.6$. Parameter $J=1.0$. Time is in units of $\hbar/J$.}
    \label{fig8}
\end{figure*}

As in the previous sections, we take the system initialized in an atomic coherent state $\ket{(\phi_0, z_0)}$. In the absence of interactions, the time evolution of such a state is perfectly coherent and periodic, and no DQPT occurs. Consider instead $\Lambda \neq 0$ and $\Pi = 0$, corresponding to the standard interacting atomic Josephson junction. When the semiclassical trajectory with initial condition $(\phi_0, z_0)$ lies well within either the Josephson or the self-trapping regime---namely, sufficiently far from a separatrix or the boundary of the libration region of a fixed point---we find clear numerical evidence that DQPTs arise for any nonzero $\Lambda$, provided the evolution time is sufficiently long. Following their first occurrence, the critical times recur in an almost periodic manner, at least until the time-evolved wave packet has lost coherence to such an extent that the interpretation of the numerical data becomes ambiguous. The value of the first critical time $t_c$ decreases with increasing $|\Lambda|$ and, for fixed $\Lambda$, lies between two semiclassical periods $T=T(\mathcal E(\phi_0, z_0))$, i.e. $nT < t_c < (n+1)T$ for some integer $n \ge 0$. The inclusion of pair tunneling substantially reshapes the distribution of complex Loschmidt zeros and shifts the critical times; however, it does not fundamentally alters the nature of the DQPTs, which continue to manifest as cusps in the return rate. Multimedia files illustrating these results for varying $\Lambda$ and $\Pi$ are provided as Supplemental Material. 

A representative example is shown in Fig.~\ref{fig8}. We see that as $N$ increases, few-body oscillations are suppressed, resulting in a smoother $\lambda(t)$ with better-defined cusp precursors, and the Loschmidt zeros tend to coalesce along well-defined lines in the complex plane; their intersections with the imaginary $\tau$ axis coincide precisely with the incipient cusps observed in the return rate. In panels (b) and (d), we observe that the introduction of pair tunneling shifts the first cluster of zeros towards the imaginary $\tau$ axis, making the first DQPT occur at a time $t_c<T$; on the other hand, the period between two consecutive DQPTs is larger than in the absence of pair tunneling. A further distinction is that, in the presence of pair tunneling, DQPTs occur significantly closer to the semiclassical return times, so that critical times deviate markedly from $\left(n + \tfrac{1}{2}\right)T$. This is related to the fact, visible in panel (d), that pair tunneling mainly affects the location of the Loschmidt zeros with larger $\mathrm{Im}\,\tau$ within each cluster.

These features admit a compelling geometrical interpretation \cite{Lang}. We can think of the large-$N$ atomic coherent state $\ket{(\phi_0, z_0)}$ as represented by a semiclassical amplitude $\Psi_0(\phi,z)$ in phase space, localized around $(\phi_0, z_0)$. Under the classical Hamiltonian flow $\Theta_t:(\phi,z) \mapsto (\phi(t), z(t))$ each point of the wave packet moves along its own trajectory, so that at time $t$ the wave packet becomes a stretched and distorted cloud, $\Psi_t(\phi,z)=\Psi_0(\Theta_{-t}(\phi,z))$, whose center of mass follows the classical trajectory starting from $(\phi_0, z_0)$. As shown in Appendix \ref{appB}, within this description the return rate at time $t$ is dominated by the phase-space point (call it the saddle point) that minimizes the arithmetic mean of its distances from the original center $(\phi_0, z_0)$ at times 0 and $t$. A DQPT occurs when the identity of such saddle point changes discontinuously at a critical time $t=t_c$. At time zero the saddle point coincides with $(\phi_0, z_0)$, and if the wave packet evolves almost rigidly the identity of the saddle point will change continuously. By contrast, interactions smear the wave packet, creating regions where different points compete to minimize the mean distance. Stronger interactions accelerate the smearing, opening the possibility for the saddle-point switch to happen earlier. 

This picture also explains why the saddle-point switch occurs between two semiclassical periods, e.g. for $T<t_c<2T$ in the case of the first DQPT of Fig.~\ref{fig8}(a). Let us imagine following the evolution of the semiclassical amplitude $\Psi_0(\phi,z)$. For $0 \le t \lesssim T$, as the wave packet stretches the saddle point remains near the evolving center of mass. For $t \gtrsim T$, as the center of mass moves away from $(\phi_0, z_0)$, the saddle point shifts gradually toward the back tail of the distorted wave packet, to remain close to the original center. As $t$ grows further, the leading edge of the wave packet swings back toward the original center (anticipating the center of mass), and at $t=t_c < 2T$ the saddle point jumps abruptly from the back to the front of the wave packet, realizing the DQPT. In this way, DQPTs can be visualized as a dramatic reorganization of the wave packet’s dominant contribution in phase space: the system switches which part of the distorted wave packet counts most for the return rate, producing the characteristic cusps in $\lambda(t)$.

\section{Discussion and conclusions}\label{sec:conclusion}

In this paper, we have studied the effects of dipolar interactions on zero-temperature static and dynamical properties of a Bose gas confined in a double-well potential, thereby realizing a dipolar atomic Josephson junction within an extended two-site Bose-Hubbard framework. Dipolar interactions are responsible for a pair tunneling term, usually neglected in the non-polarized case, which significantly alters both the equilibrium and nonequilibrium properties of the system.

In equilibrium, we have shown that pair tunneling substantially modifies the structure of the many-body ground state. In particular, it induces parity modulations in the distribution of Fock-state probabilities and leads to a broader population spread across Fock states, reflecting enhanced many-body correlations. Concerning quantum criticality, we demonstrated that the quantum phase transition toward the NOON state---corresponding to an imbalance-fragmented condensate---is strongly affected by pair tunneling. In general, it shifts the critical interaction strength and, above a threshold, qualitatively changes the nature of the transition from continuous to first order. Furthermore, pair tunneling gives rise to a distinct continuous quantum phase transition toward a phase-NOON state, which, like the imbalanced NOON state, represents a fragmented condensate. 

At the dynamical level, we showed that the familiar dynamical regimes---Josephson oscillations and macroscopic quantum self-trapping---persist in the presence of pair tunneling. However, the modified semiclassical structure introduces additional fixed points in phase space, leading to new dynamical features. These include phase-imbalanced Josephson oscillations with $\langle \phi(t)\rangle_t \neq \{0,\pi\}$, oscillations around multiple minima, and modified conditions for the onset of macroscopic quantum self-trapping. Thus, while the overall dynamical taxonomy remains intact, the underlying phase-space organization is significantly enriched. Finally, we analyzed dynamical quantum phase transitions, showing that several features of the exact numerical results admit a clear geometric interpretation in terms of the evolution and deformation of semiclassical phase-space amplitudes. Pair tunneling induces quantitative modifications in the DQPT structure, both the critical times and their periodic spacing, but does not lead to a fundamentally different type of dynamical critical behavior. 

In general, the two-mode effective model \eqref{Hamiltonian} on which our analysis is based is justified in the low-energy, weak-overlap regime of a sufficiently deep double-well potential, where the condensate remains predominantly localized in the two lowest single-particle orbitals and the mixing of higher modes is negligible. Within this framework, it is important to distinguish between two levels of description: the semiclassical two-mode dynamics \eqref{semiclas}, and the fully quantum dynamics of the effective Bose-Hubbard model \eqref{Hamiltonian}, which additionally captures number fluctuations and many-body correlations within the restricted two-mode subspace \cite{Milburn}. 

For dipolar condensates, previous studies have benchmarked semiclassical two-mode predictions against full three-dimensional Gross-Pitaevskii simulations, showing that the effective description successfully reproduces the essential Josephson and self-trapping phenomenology, with quantitative agreement improving when the Josephson frequency is small compared to the characteristic trap frequencies \cite{Blume, Adhikari}. We emphasize, however, that the models employed in those works were simpler than the one considered here, as they neglected the pair-tunneling contribution responsible for the $\cos(2\phi)$ term in our classical Hamiltonian. Our treatment therefore extends the standard two-mode framework by incorporating additional interaction-induced tunneling processes. Moreover, in the case of contact interactions, it was demonstrated that the quantum two-mode Bose-Hubbard model provides a significantly improved description over the semiclassical approximation, yielding better agreement with exact many-body calculations \cite{Sakmann}. Although an analogous benchmark has not been carried out explicitly for dipolar systems, there is strong reason to expect that the inclusion of full quantum dynamics similarly enhances the reliability of the effective description in the present case. At the same time, the model remains inherently limited: phenomena such as roton-like excitations, transverse instabilities, and collapse mechanisms \cite{Lewenstein_book} lie beyond the scope of any two-mode treatment and require fully three-dimensional mean-field or many-body approaches for a complete description. These effects, however, fall outside the low-energy regime considered here, under which we expect our conclusions to remain qualitatively robust.

Within this regime, the effective two-site model is characterized by the renormalized interaction and tunneling parameters ${U = U_0 - V}$ and ${J = J_0 + T(N-1)}$, leading, in the experimentally relevant large-$N$ limit, to the dimensionless parameters $\Lambda = UN/2J = {(U_0-V)N}/{[2J_0+2T(N-1)]} \simeq (U_0-V)/2T$ and $\Pi = PN/J=PN/[J_0+T(N-1)] \simeq P/T$. Since $U_0$ can be tuned over a broad range through standard scattering-length control, $\Lambda$ may attain both large positive and negative values, or be brought close to zero when $U_0 \simeq V$. Similarly, depending on the double-well geometry and the applied electric or magnetic fields, the ratio $P/T$ can vary over the range $\Pi \sim 0.01-1$. These parameter regimes fully encompass those required to observe the phenomena predicted in this work. We therefore expect that the dipolar many-body effects discussed here should be accessible with current experimental platforms, without posing challenges beyond those routinely encountered in the field.

Overall, our results demonstrate that dipolar interactions provide a controlled mechanism for engineering higher-order tunneling processes in atomic Josephson junctions, enabling significant modifications of both equilibrium and dynamical properties and of the associated quantum critical phenomena. In view of the ongoing interest in Josephson dynamics both as a fundamental phenomenon and in extended and dipolar quantum systems, especially in the context of supersolids, we expect our results to provide useful insight into higher-order tunneling processes and their impact on equilibrium and dynamical criticality in these emerging platforms.

\section*{Acknowledgments}

C.V. thanks Marco Di Liberto and Qizhong Zhu for useful discussions. This work is partially supported by the Project ``Frontiere Quantistiche'' (Dipartimenti di Eccellenza) of the
Italian Ministry of University and Research (MUR), by ``Iniziativa Specifica Quantum'' of INFN, by the European Union-Next Generation EU within the European Quantum Flagship Project ``PASQuanS 2'', the National Center for HPC, Big Data and Quantum Computing (Project No.~CN00000013, CN1 Spoke 10: ``Quantum Computing''), and by the PRIN Project “Quantum Atomic Mixtures: Droplets, Topological Structures, and Vortices” of MUR.

\appendix

\section{Mean-field phase diagram}\label{app:fig}

\begin{figure*}
    \centering
    \includegraphics[width=\linewidth]{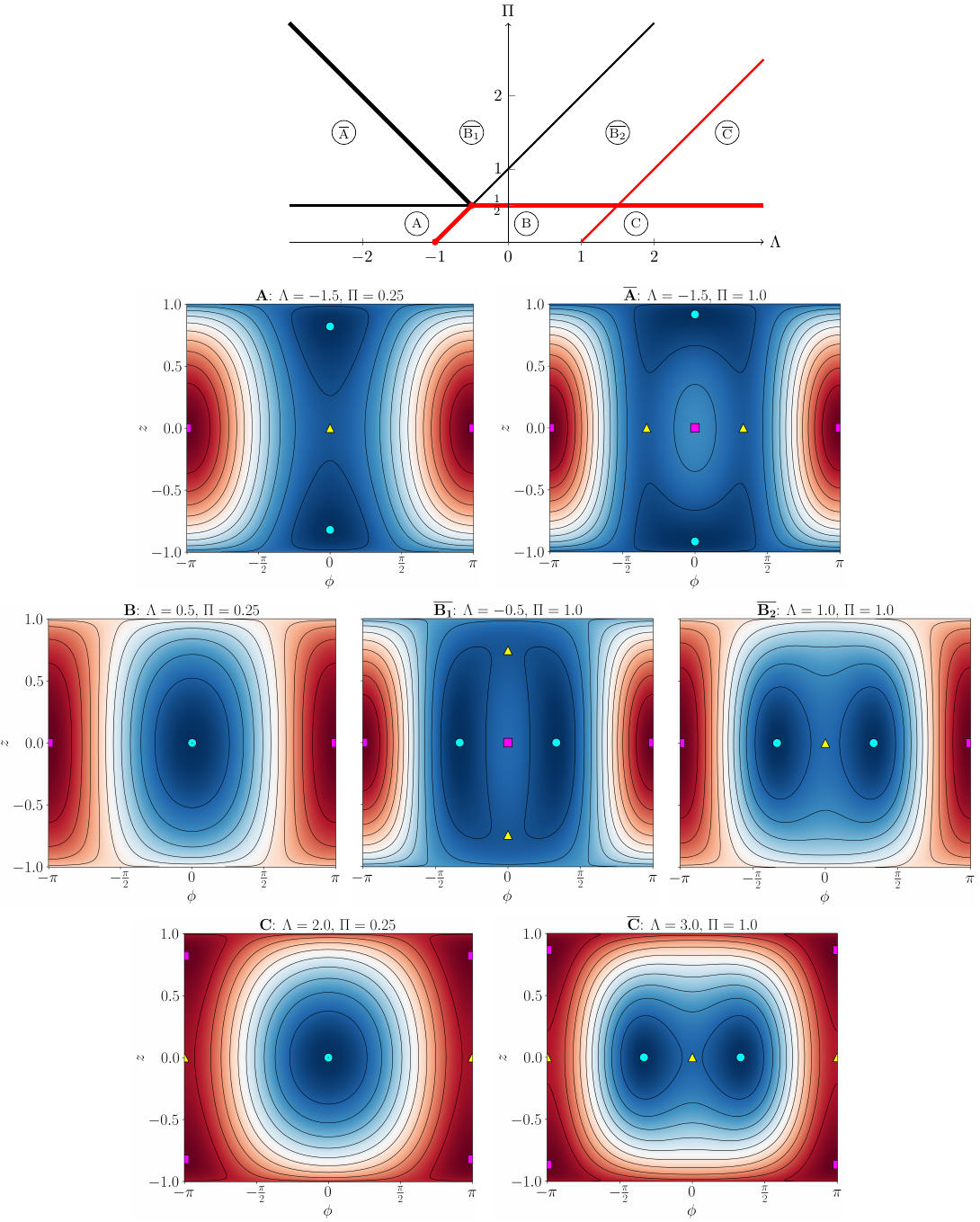}
    \caption{\textbf{Mean-field phase diagram (extended version).}}
    \label{fig9}
\end{figure*}

In Fig.~\ref{fig9}, we present an extended version of the mean-field phase diagram discussed in Sec.~\ref{sec:qpt}. For each of the seven regions identified in Fig.~\ref{fig2} and Table \ref{tab1}, characterized by distinct configurations of stationary points, we show a representative contour plot of $\mathcal E(\phi,z)$ in phase space. Each contour represents a possible classical orbit. Dark blue indicates lower values of $\mathcal E$, while dark red indicates higher values. Local minima are marked by cyan circles, local maxima by magenta squares, and saddle points by yellow triangles.

\section{Loschmidt cumulants}\label{appA}

The method of Loschmidt cumulants \cite{Peotta} provides an efficient framework for the accurate determination of the complex zeros of the Loschmidt amplitude in the vicinity of a given basepoint $\tau$. Here we give a brief outline of the method, referring the reader to Appendices A and B of Ref.~\cite{Peotta} for additional details.

One defines Loschmidt moments 
\begin{equation}\label{moment}
    \langle \hat H^n\rangle_\tau = (-\hbar)^n\frac{\partial^n_\tau \mathcal Z(\tau)}{\mathcal Z(\tau)} = \frac{\braop{\Psi_0}{\hat H^n}{\Psi(\tau)}}{\braket{\Psi_0}{\Psi(\tau)}}
\end{equation}
and cumulants
\begin{equation}
    \llangle \hat H^n \rrangle_\tau = (-\hbar)^n \partial^n_\tau \ln \mathcal Z(\tau),
\end{equation}
which are related recursively via
\begin{equation}\label{cumulant}
    \llangle \hat H^n \rrangle_\tau = \langle \hat H^n\rangle_\tau - \sum_{m=1}^{n-1} \binom{n-1}{m-1}\llangle \hat H^m \rrangle_\tau \langle \hat H^{n-m}\rangle_\tau,
\end{equation}
with $\llangle \hat H \rrangle_\tau = \langle \hat H \rangle_\tau$. Using Eq.~\eqref{Losch_zeros}, we can express properly normalized cumulants $\kappa_n(\tau)$ in terms of simple summations over Loschmidt zeros,
\begin{equation}
    \kappa_n(\tau) = \frac{(-1)^{n-1}}{\hbar^n (n-1)!}\llangle \hat H^n\rrangle_\tau = \sum_j \frac{1}{(\tau_j-\tau)^n},
\end{equation}
with $n>1$. Therefore the cumulants are primarily determined by the Loschmidt zeros located nearest to the complex basepoint $\tau$, whereas the influence of more distant zeros is suppressed by the inverse of their distance to the basepoint raised to the power of the cumulant order. This allows to make the problem tractable by truncating the summation on the right-hand side to the $m$ distinct zeros closest to the basepoint, each counted with its multiplicity $d_j$:
\begin{equation}\label{approxcum}
    \kappa_n(\tau) \simeq \sum_{j=1}^m d_j\lambda^n_j,
\end{equation}
where $\lambda_j \equiv (\tau_j-\tau)^{-1}$. The approximation becomes increasingly accurate at higher cumulant orders $n$. 

In this truncated form, the cumulants satisfy a homogeneous linear difference equation of degree $m$ for some coefficients $a_l$,
\begin{equation}
    \kappa_n = a_1 \kappa_{n-1} + a_2\kappa_{n-2} + \cdots + a_m\kappa_{n-m},
\end{equation}
and the numbers $\lambda_j$ are the $m$ solutions of the associated characteristic equation. The first step is thus to compute the coefficients $a_l$ by solving a linear system of $m$ equations. This requires the knowledge of $2m$ cumulants $\kappa_l$ for $l = n-m,\,\dots,\,n+m-1$, which we can compute exactly according to Eqs.~\eqref{moment} and \eqref{cumulant}. The second step consists of solving the characteristic equation
\begin{equation}
    \lambda^m-a_1\lambda^{m-1}-a_{m-1}\lambda-a_m=0
\end{equation}
to determine the $m$ roots $\lambda_j$. The approximate Loschmidt zeros are then obtained as $\tau_j = \tau + 1/\lambda_j$. The last step is to compute the multiplicities $d_j$. Since Eq.~\eqref{approxcum} is an approximation, they are in general non-integer. We therefore identify the most reliable approximate zeros by retaining only the pairs $(\tau_j, d_j)$ for which $|d_j-\ell|<0.01$, where $\ell$ is an integer. In our case we have found no zeros with multiplicity $\ell > 1$. The procedure is then repeated while varying the basepoint $\tau$ along the imaginary axis within the time interval of interest.

The results in panels (c) and (d) of Fig.~\ref{fig8} are obtained following this procedure for $m=10$ and $n=50$, with 200 basepoints along the imaginary $\tau$ axis from $t=0$ to $t=6.0$.

\section{Geometrical picture of DQPTs}\label{appB}

Let $\bm v=(\phi, z)$ denote a point in classical phase space and $\ket{\bm v}$ the Glauber coherent state parametrized by $\bm v$, normalized such that $\int d\bm v\,\ket{\bm v}\bra{\bm v}= \hat 1$. The large-$N$ system is initialized in an atomic coherent state centered around $\bm v_0$, $\ket{\Psi_0} = \int d\bm v\,\Psi_0(\bm v)\ket{\bm v}$, where $\Psi_0(\bm v) \propto e^{-\frac{N}{4}[\mathcal A_0(\bm v)+\iunit \mathcal S_0(\bm v)]}$ and $\mathcal A_0$ and $\mathcal S_0$ are the real and imaginary parts of a phase-space rate function: $\mathcal A_0(\bm v)$ quantifies the phase-space distance from $\bm v_0$, while $\mathcal S_0(\bm v)$ encodes the local geometric phase. Using the coherent state path integral, the Loschmidt amplitude can thus be written as
\begin{align}
    \mathcal Z(t) &= \braop{\Psi_0}{e^{-\iunit \hat H t/\hbar}}{\Psi_0}\nonumber\\
    &= \int d\bm v\,d\bm v'\,\Psi_0^*(\bm v') \Psi_0(\bm v) \braop{\bm v'}{e^{-\iunit \hat H t/\hbar}}{\bm v}\nonumber\\
    &= \int d\bm v\,d\bm v'\,e^{-\frac{N}{4}[\mathcal A_0(\bm v) + \mathcal A_0(\bm v') + \iunit (\mathcal S_0(\bm v) - \mathcal S_0(\bm v'))]}\nonumber\\
    &\quad \times \int\limits_{\bm u(0)=\bm v}^{\bm u(t)=\bm v'} \mathcal D\bm u\,e^{\frac{\iunit N}{\hbar}s(\bm u)},
\end{align}
At large $N$, the path integral is evaluated by stationary phase, which selects $\bm v'$ as $\bm v(t)$, i.e. the classical time evolution of $\bm v$. Therefore
\begin{equation}
    \mathcal Z(t) \simeq \int d\bm v\,e^{-\frac{N}{4}\left[\mathcal A_0(\bm v) + \mathcal A_0(\bm v(t)) - i(\frac{4 s_\text{cl}}{\hbar}+\mathcal S_0(\bm v(t))-\mathcal S_0(\bm v))\right]},
\end{equation}
where $s_\text{cl}$ is the action \eqref{action} evaluated on the classical trajectory $\bm v \mapsto \bm v(t)$. Taking the modulus square, the oscillatory phase drops out at leading exponential order, yielding
\begin{equation}
    |\mathcal Z(t)|^2 \simeq \int d\bm v\,\exp\left[-N\frac{\mathcal A_0(\bm v(t)) + \mathcal A_0(\bm v)}{2}\right].
\end{equation}
This integral is dominated by the saddle point $\overline{\bm v}_t$ that minimizes $[\mathcal A_0(\bm v(t)) + \mathcal A_0(\bm v)]/2$. Geometrically, this corresponds to selecting the classical trajectory whose initial and time-evolved phase-space points minimize the arithmetic mean of their distances from the initial center $\bm v_0$. The return rate is therefore
\begin{equation}
    \lambda(t) \simeq \min_{\bm v} \frac{\mathcal A_0(\bm v(t))+\mathcal A_0(\bm v)}{2},
\end{equation}
and a DQPT occurs when the minimizing vector $\overline{\bm v}_t$ jumps discontinuously at a critical time $t=t_c$, leading to a nonanalyticity (cusp) in $\lambda(t_c)$.

\bibliography{References}

\end{document}